\documentclass[aps,twocolumn,superscriptaddress,altaffilletter,lengthcheck, tightenlines,showpacs,showkeys]{revtex4}

\newcommand{\al}{\alpha}

\newcommand{\D}{\Delta}
\newcommand{\ben}{\begin{eqnarray}}
\newcommand{\een}{\end{eqnarray}}
\newcommand{\be}{\begin{equation}}
\newcommand{\ee}{\end{equation}}
\newcommand{\ba}{\begin{eqnarray}}
\newcommand{\ea}{\end{eqnarray}}
\newcommand{\n}{\label}
\newcommand{\no}{\noindent}

\newcommand{\ga}{\gamma}
\newcommand{\ro}{\rho}
\newcommand{\om}{\omega}

\newcommand{\bn}{\begin{equation}\label}

\usepackage[dvipdf]{epsfig}
\usepackage{color}

\begin{document}

\title{Interacting realization of cosmological singularities with variable vacuum energy}

\author{Luis P. Chimento}%\email{chimento@df.uba.ar}
\affiliation{Departamento de F\'{\i}sica, Facultad de Ciencias Exactas y Naturales,  Universidad de Buenos Aires and IFIBA, CONICET, Ciudad Universitaria, Pabell\'on I, Buenos Aires 1428 , Argentina}
\author{Mart\'{\i}n G. Richarte}%\email{martin@df.uba.ar}
\affiliation{Departamento de F\'{\i}sica, Facultad de Ciencias Exactas y Naturales,  Universidad de Buenos Aires and IFIBA, CONICET, Ciudad Universitaria, Pabell\'on I, Buenos Aires 1428 , Argentina}

\bibliographystyle{plain}

\begin{abstract}
We examine  an interacting dark matter--variable vacuum energy  model  for a spatially flat Friedmann-Roberston-Walker spacetime, focusing on  the appearance of cosmological singularities  such as \emph{big rip, big brake,  big freeze}, and \emph{ big separation}  along with abrupt events (\emph{infinite $\gamma$- singularity} and \emph{new w-singularity}) at late times.  We introduce a phenomenological interaction which has a nonlinear dependence on the total energy density of the dark sector and its derivative, solve exactly the source equation for the model and find the energy density as function of the scale factor as well as  the time dependence of the approximate scale factor in the neighborhood of the singularities.  We describe the main characteristics  of these singularities by  exploring the  type of interaction that makes them possible along with behavior of dark components  near them.  We  apply the geometric  Tipler  and Kr\'olak  method for determining the fate of time-like geodesic curves around the singularities.  We also explore the strength  of them by  analyzing the leading term in  some geometric invariants  such as the square Riemann scalar and the Ricci scalar.  \\

 \end{abstract}
\vskip 1cm


\date{\today}
\maketitle
%%%%%%%%%%%%%%%%%%%%%%%%%%%%%%%%%%%%%%
\section{Introduction}
%%%%%%%%%%%%%%%%%%%%%%%%%%%%%%%%%%%%%
In 1998 astrophysical observations coming from distant supernovae--stellar explosions led to the  inevitable conclusion that our Universe is actually speeding up rather than  slowing down  \cite{Book}  due to the existence of a repulsive  agent  known  as dark energy.  Such outstanding finding  was  promptly confirmed  by 
additional cosmological observations based on the measurements of   cosmic microwave background anisotropies, baryon acoustic oscillations, and  power spectrum of clustered matter \cite{Planck2013, WMAP9}. Despite the plethora of observational evidences, in favor of  the present  acceleration of the Universe,  accumulated  over the last years  by studying even more remote supernovae and launching new satellites \cite{Planck2013}, quite little is known about the true nature of dark energy. For instance,  there is not a fundamental theory   which can account for  the origin of dark energy  at microscopic level.

In order to shed some light over the dark side of the Universe  one could devote some efforts to explore  its ultimate fate  \cite{tipo1}. An appealing route to follow is to analyze some  dark energy models which can explain the current speeding up of the Universe but  exhibit the presence of  a cosmic singularity  in the asymptotic (remote) future \cite{tipo1}, \cite{tipo1b}; so one may link the present state of the Universe with its drastic final state.  A fundamental task  to be addressed by these alternative scenarios is the classification of  singularities that could emerge at a finite time or abrupt events \cite{tipo1}, \cite{tipo1b}, \cite{tipo2}, \cite{tipo2a}, \cite{tipo2aa}, \cite{tipo2b}, \cite{tipo3}, \cite{mariam}, \cite{mariam2}, \cite{tipo4}.  Several  useful ways to characterize such final doomsdays  are based on geometric methods  which examine the existence of causal geodesics that cannot be extended to arbitrary values of their proper time (geodesic incompleteness) \cite{haw} or  the possibility to show that  geodesic curves can be extended beyond cosmic singularity \cite{ruth}, \cite{barrow}. Another manner consists in taking into account  the behavior of curvature invariants near the singularity  so the strength of singularities can be determined using the necessary and sufficient conditions obtained by   Tipler \cite{tipler} and Kr\'olak \cite{krolak}. These conditions are considerably  useful provided  help to classify singularities as strong and weak types, giving some insights on the magnitude of the tidal forces experienced by a co-moving observer toward the singularity. In fact,   the Tipler definition requires that any object  has its volume crushed to zero  as the singularity is approached  \cite{tipler} whereas  the Kr\'olak definition is weaker than the Tipler version  and is related  with inquiries on  the cosmic  censorship conjecture \cite{k2}. 

A serious approach  for understanding  the nature of cosmological singularities requires to  take into account the behavior of dynamical variables which enter into the field equations. Therefore  the blow up of  the energy density and  the divergent behavior of the  pressure  seem to be of interest along with the behavior of geometric quantities such as the scale factor, the Hubble function and its derivatives. In this way,  one must focus on  certain physical properties  which  make these kinds of singularities fairly distinctive among them and how such traits can determine the ultimate fate of the universe \cite{tipo1b}, \cite{tipo2}, \cite{tipo2a}, \cite{tipo2aa}, \cite{tipo2b}, \cite{tipo3}, \cite{mariam}, \cite{mariam2}, \cite{tipo4}. At this point,  it would seem  crucial to explore a viable cosmological scenario where  the aforesaid singularities
can appear naturally in order to explore its physical outcome.  To do that,  we are going to present some interacting dark energy models \cite{jefe1}, \cite{jefe2} with the presence of singularities (or singular event) and abrupt events.  A natural question that could arise is what kind of phenomenological interactions do imply the existence of a future singularity or abrupt event.   We must stress that our approach  for studying cosmic singularities is considerably different from previous works, thus, we focus in an interacting dark energy model  with a  nonlinear interaction term and then analyze the ultimate fate of the universe.  As a consequence of this method, we will firstly study the existence of singularity along with its traits without imposing any particular type of scale factor.  However, we will  \emph{obtain}  the approximate scale factor or the exact one, in the cases where it is  possible, to illustrate the model. Since we do not postulate or give the form of  scale factor from the very beginning, it is clear that  only some singularities will emerge in our model. In this way, we will construct  \emph{a posteriori} a scale factor associated with a cosmic singularity for a given interaction. Such expression will be useful for  studying  then  the behavior of  the Hubble function, its derivatives in addition to the behavior of  the energy density and  pressure in terms of the cosmic time.

Our goal is  then to  consider an interacting dark matter--variable vacuum energy (VVE) framework  and   explore  the new details coming from the appearance of cosmic singularities in the remote future (distant past) or abrupt/singular events. Taking into account that these viable models exhibit a current accelerating phase together with  a future doomsday, we are going to describe the physical traits associated with them.  Further,  we analyze several kinds of singularities, make some comments on the type of interaction supporting them, in particular, we focus on the behavior of both dark component  near the singularity. Finally, we examine these singularities  with the help of the  Tipler  and Kr\'olak criteria.

%%%%%%%%%%%%%%%%%%%%%%%%%%%%%%%%%%%%%%%%%%%%%%%%%%%%%% 
\section{Interaction framework}
%\section{Description of the model}
%%%%%%%%%%%%%%%%%%%%%%%%%%%%%%%%%%%%%%%%%%%%%%%%%%%%%%

 To study  the interacting dark sector model we consider the  spatially flat Friedmann-Roberston-Walker (FRW) metric $ds^2=-dt^2+a^2(t)dx^idx_i$, where $a(t)$ is the scale factor and $dx^{i}dx_{i}$ is the line element corresponding to hypersurface of constant time. The universe is filled with two perfect fluids, one  accommodates as a matter component  while the the other represents  a VVE substratum. Both  fluids  are described by linear equations of state, having energy densities $\ro_m$, $\ro_x$ and pressures $p_m$, $p_x$,  respectively. The total energy density $\ro$  and the conservation equation $\dot\rho+3H(\rho+p)=0$ for this interacting two-fluid model can be written as 
\bn{aa}
\ro=\ro_m+\ro_x,
\ee
\bn{co}
\ro'=-\ga_m\ro_m-\ga_x\ro_x,
\ee
where the dot  stands for derivative with respect to the cosmic time $\dot{} \equiv d/dt $ being $H=\dot a/a$, the  prime is the derivative with respect to scale factor $'\equiv d/d\eta=d/3Hdt=d/d\ln{(a/a_0)^3}$ and $a_0$ is some value of reference for the scale factor.  We have assumed  that both interacting components admit an equations of state $p_i=(\ga_i-1)\ro_i$  with $i=\{m, x\}$ such that the constant barotropic indexes $\ga_m$ and $\ga_x$ satisfy the next condition:  $0<\ga_x<\ga_m$. After having solved the algebraic linear system of  equations (\ref{aa})-(\ref{co}),  we are able to get $\ro_m$ and $\ro_x$ as functions of $\ro$ and its $\eta$ derivative $\ro'$: 
\be
\n{rom}
\ro_m=-\,\frac{\ga_x\ro+\ro'}{\ga_m-\ga_x}, 
\ee
\bn{rox}
\ro_x=\frac{\ga_m\ro+\ro'}{\ga_m-\ga_x}.
\ee
To complete the model we introduce an exchange of energy in the dark sector  in terms of a factorized interaction $3HQ(\ro,\ro',\eta)$. Concerning this aim, we split the balance equation (\ref{co}) as follows: 
\bn{qm}
\ro'_m+\ga_m\ro_m=-Q,   
\ee
\bn{qx}
\ro'_x+\ga_x\ro_x=Q.
\ee
From Eqs. (\ref{rom})-(\ref{qx}), we arrive at  a second order differential equation for the energy density  
\be
\n{se} 
\rho''+(\ga_m+\ga_x)\rho'+\ga_m\ga_x\rho= Q(\ga_m-\ga_x),
\ee
that  we will call ``source equation" \cite{jefe1} henceforth.

The uniqueness of the solutions (\ref{rom})-(\ref{rox})  corresponding to  the  algebraic system equations (\ref{aa})-(\ref{co}) allows us to extract some interesting conclusions of the model. In the procedure outlined above one can  find the  energy density $\ro$ by solving the source equation (\ref{se}) for a given interaction term $Q$, subsequently,  the energy densities of the matter and VVE components are reconstructed by means of Eqs. (\ref{rom})-(\ref{rox}), pointing that this procedure does not rely on the specific cosmological equations that govern the dynamic of an homogeneous isotropic flat universe. Later on, we will investigate a concrete example for a given interaction term and obtain several general features about the existence of initial and final singularities in the interacting dark sector model without  knowing  of the scale factor. 

As a result of this approach,  we have been reducing the interacting framework to an effective one-fluid model with energy density $\ro=\ro_m+\ro_x$ and  total pressure $p(\ro,\ro')=p_m+p_x=-\ro-\ro'$. Comparing the later equation with the effective equation of state of the dark sector $p=(\gamma-1)\rho$ , we obtain its effective conservation equation
\bn{ce} 
\ro'+\ga\ro=0, 
\ee
where the effective barotropic index reads 
\bn{ge}
\gamma=\frac{(\ga_m\ro_m+\ga_x\ro_x)}{\rho}. 
\ee

In calculating  the scale factor of the homogeneous and isotropic flat universe, we adopt the  Einstein field equations, so that the dynamic of the effective one-fluid model will be governed by the corresponding Friedmann constraint,
\bn{00}
3H^2=\rho.
\ee

We will investigate  a universe which transits from an initial matter-dominated phase into a  final era dominated  by an unknown component  that will be identified with VVE, the former component is associated with an initial singularity and the latter one with a final singularity or a possible  doomsday of the universe. In our model the VVE has an equation of state of the form $p_x=-\ro_x$, so its barotropic index vanishes ($\ga_x=0$), and therefore the source equation (\ref{se}), the matter energy density (\ref{rom}), and the VVE density (\ref{rox})  turn to be  given by 
\bn{sef} 
\rho''+\ga_m\rho'= \ga_m Q,
\ee
\bn{rmxf}
\ro_m=-\frac{\ro'}{\ga_m}, \qquad \ro_x=\ro+\frac{\ro'}{\ga_m}.
\ee
So far we have been speaking of  an exchange of energy between matter and VVE fluids without specifying its form. We now propose  that the interaction term in the dark sector can be  defined  via the equation  
\be
\n{q}
Q=-\frac{df}{d\eta}.
\ee
Here  we suppose that the input function $f$ only  depends on $\ro, \ro'$, and $\eta$ for simplicity.  Replacing this interaction term $Q$ into the source equation (\ref{se}), it leads us  to a nonlinear differential equation 
\be
\n{x''}
\ro''+\ga_m \ro'=-\ga_m f'.
\ee
Taking into account the  first integral of Eq. (\ref{x''}),  it allows us to write the equation of state of the mixture $p=-\ro-\ro'$,  the effective barotropic index $\ga=-\ro'/\ro$, the acceleration of the universe $\ddot a$ along with  the matter and VVE densities (\ref{rmxf}) as follows 
\bn{ro'}
\ro'=-\ga_m(c+\ro+f),
\ee 
\bn{p}
p=(\ga_{m}-1)\rho+ \ga_{m}(c+f),
\ee
\bn{ga}
\ga=\ga_m\left[1+\frac{c}{\ro}+\frac{f}{\ro}\right],
\ee
\bn{a..}
\frac{\ddot a}{a}=-\frac{1}{6}(3\ga_m-2)\ro-\frac{\ga_m}{2}(c+f),
\ee
\bn{rmx,f}
\ro_m=c+\ro+f, \qquad \ro_x=-(c+f).
\ee
where $c$ is an integration constant. From Eqs. (\ref{ro'})-(\ref{rmx,f}), we observe that the $\eta$ derivative of the energy density, the pressure, the effective barotropic index, the acceleration term, the matter and VVE densities, all of them depend linearly  with the input function $f$ to describe the whole contribution coming from the energy transfer in the dark sector (\ref{q}).  An interesting point in regard with  the role played by the interaction can be understood by analyzing a regime where  $f$ gives  the largest contribution, neglecting  $\rho$ and $c$. In such regime, the previous dynamical quantities can be easily found  as follows 
$$
\ro'\approx-\ga_m f,~~p\approx\ga_m f,~~ \ga\approx\ga_m f/\ro, 
$$
\bn{intapp}
\ddot a\approx-\ga_m af/2, \quad \ro_m\approx-\ro_x\approx-f.
\ee
\no In order to further motivate our results, let us assume that function $f(t)\to f(t_s)$ when  $t\to t_s$ so that in the limit case  $f(t_s)=\pm\infty$, blowing up  at the finite value of the cosmic time $t_s$. Then, we have that the cosmic time derivative of the Hubble variable $\dot H=\ro'/2\approx-\ga_m f/2\to\mp\infty$, the pressure $p\approx\ga_m f\to\pm\infty$,  the effective barotropic index $\ga\approx\ga_m f/\ro\to\pm\infty$, the acceleration $\ddot a\approx-\ga_m af/2\to\mp\infty$ and the matter and VVE densities $\ro_m\approx f\to\pm\infty$, $\ro_x\approx-f\to\mp\infty$, indicating that all these quantities diverge as $t\to t_s$.  This promising approach then  opens the possibility of producing an initial or final  singularity at $t_s$  within the framework of  interacting  dark sector, so it will be useful for examining the ultimate fate of a universe. In the next section, we are going to generate $Q$ by selecting the input function $f$ and examine its physical outcome.

%%%%%%%%%%%%%%%%%%%%%%%%%%%%%%%%%%%%%%%%%%%%%%%%%%%%%%%%%%%%%%%%%%%%%
\section{Interaction framework produces initial and final singularities}
%%%%%%%%%%%%%%%%%%%%%%%%%%%%%%%%%%%%%%%%%%%%%%%%%%%%%%%%%%%%%%%%%%%%%

%%%%%%%%%%%%%%%%%%%%%%%%%%%%%%%%%%%%%%%%%%%%%%%%%%%%%%%%%%%%%%%%%%%%%
\subsection{General properties}
%%%%%%%%%%%%%%%%%%%%%%%%%%%%%%%%%%%%%%%%%%%%%%%%%%%%%%%%%%%%%%%%%%%%%

 In this section, we wish to investigate the initial and final singularities of the universe when the interaction in the dark sector is generated  by the function $f$, written as  a power-law of the energy density  
\bn{f}
f=\al\ro^{-n},
\ee
which yields an exchange of energy  associated with the nonlinear interaction term (\ref{q})
\bn{q1} 
Q=n\al\ro^{-n-1}\ro',
\ee
where $\al$ is a coupling constant and $n$ is a non-vanishing  real number. Then the first integral (\ref{ro'}) becomes
\bn{ro'1}
\ro'=-\ga_m(c+\ro+\al\ro^{-n}).
\ee 
To simplify our formulation, we  slightly  redefine the interaction (\ref{q1}) to get  a scenario where the integration constant $c$ vanishes and does not contribute to the model.  We  insert the first integral (\ref{ro'1}) with $c=0$ into Eq. (\ref{q1}) and  we obtain the final interaction 
\bn{qf}
Q=-n\al\ga_m \ro^{-n-1}\left(\ro+\al\ro^{-n}\right),
\ee
which  can also be written as  
\bn{qq}
 Q=-n\left(\frac{\ro'^2}{\ga_m\ro}+\ro'\right).
\ee
From  Eqs. (\ref{q}), (\ref{f}) and  (\ref{qq}), we obtain the solutions of the source equation (\ref{sef}) in the case of  $c=0$. This kind of interaction is very interesting because the effective equation of state of the effective one-fluid $p=-\ro-\ro'$ for $c=0$ turns to be that of  a Chaplygin or anti-Chaplygin gas  depending on the sign of the coupling constant $\al$, as was noticed in Ref.\cite{jefe1}. Summing up, Eqs. (\ref{ro'})-(\ref{rmx,f}) can be recast as  
\bn{ro'f}
\ro'=-\ga_m(\ro+\al\ro^{-n}),
\ee 
\bn{pcha}
p=(\ga_{m}-1)\rho+ \al\ga_{m}\rho^{-n},
\ee
\bn{gaf}
\ga=\ga_m\left[1+\al\ro^{-n-1}\right],
\ee
\bn{a..f}
\frac{\ddot a}{a}=-\frac{1}{6}(3\ga_m-2)\ro-\frac{\al\ga_m}{2}\ro^{-n},
\ee
\bn{rmxff}
\ro_m=\ro+\frac{\al}{\ro^n}, \qquad \ro_x=-\frac{\al}{\ro^n}.
\ee

We   are in condition to go a step forward for integrating  (\ref{ro'f})  and  obtaining  the energy density as a function of the scale factor
\bn{x}
\ro=\left[-\al+ b\left(\frac{a_{0}}{a}\right)^{3\ga_m(n+1)}\right]^{1/(n+1)},
\ee
where $a_0$ and $b$ are both arbitrary integration constants. A singularity can be achieved  if  any of the physical quantities, such as the energy density (\ref{x}), the pressure (\ref{pcha}), the barotropic index (\ref{gaf}) or the acceleration (\ref{a..f}), vanishes or diverges at a finite  time $t_s$. One way  to meet this  condition is by choosing the integration constant $b$ so that the square bracket in Eq. (\ref{x}) vanishes as the scale factor reaches the finite value $a_s$, namely  $b =\al (a_s/a_{0})^{3\ga_m(n+1)}$. Finally, the energy density (\ref{x}), its $\eta$ derivative and the effective barotropic index are given by 
\bn{rf}
\ro=\left\{\al\left[-1+ \left(\frac{a_s}{a}\right)^{3\ga_m(n+1)}\right]\right\}^{1/(n+1)},
\ee
\bn{r'f}
\ro'=-\al\ga_m\left(\frac{a_s}{a}\right)^{3\ga_m(n+1)}\ro^{-n},
\ee
\bn{gaa}
\ga=\frac{\ga_m}{1-\left(\frac{a}{a_s}\right)^{3\ga_m(n+1)}}\,.
\ee
To make contact with previous results we will  examine  the behavior of several  important quantities explicitly. We start with the case $\al>0$. We  achieve that the energy density $\ro\to 0$ for  $a\to a_s$  with  $n>-1$ provided  $a\le a_s$  or  in the complementary scenario where  $a\to\infty$  and  $n<-1$.  The energy density blows up ($\ro\to\infty$) under two different conditions: i-  $a\to a_s$ with $a\ge a_s$ and $n<-1$ or  ii- $a\to 0$ and $n>-1$. Another branch to study corresponds to a negative coupling constant, $\al<0$. In this case,  the energy density $\ro\to 0$ as $a\to a_s$ with $a\ge a_s$ and $n>-1$. On the contrary, the energy density becomes divergent ($\ro\to\infty$)  in the  $a\to a_s$ limit  if  two conditions simultaneously holds:  $a\le a_s$ and $n<-1$. In the last case,  , we also have that the energy densities tends to the constant  which is indeed an exact solution $\ro\to\ro_c=(-\al)^{1/(n+1)}$ [ see  Eq. (\ref{ro'f})]. It can  be obtained in the  $a\to 0$ limit for $n<-1$ or  in the case  $a\to\infty$ for $n>-1$ [see Eq. (\ref{x})]. This constant solution gives the exact de Sitter scale factor,
\bn{ds}
a_{_{dS}}=\exp{\left[\frac{(-\al)^{1/(n+1)}}{3}\right]^{1/2}\,t}.
\ee
The $\eta$ derivative $\ro'=2\dot H$ is negative or positive definite and therefore the Hubble variable is a decreasing or an increasing function of the  time, respectively. An interestingly fact is that the barotropic index always diverges ($\ga\to\infty$) in the  $a\to a_s$ limit   for any value of $n$. The remaining quantities, namely the pressure, the acceleration, and the dark matter and VVE densities can be obtained as functions of the scale factor by combining the equations (\ref{pcha}), (\ref{a..f}), (\ref{rmxff})  along with  (\ref{rf}). 

By considering the second term of the first integral (\ref{ro'f}) as the leading contribution [see Eqs. (\ref{intapp}), (\ref{f}))], it is  possible to  find the energy density after integrating the  approximate conservation equation ($\dot\ro\approx-\sqrt{3}\al\ga_m\ro^{-n+1/2}$) to arrive at  
\bn{rot} 
\ro\approx \left[-\frac{\sqrt{3}\,\al\ga_m}{2(\nu-1)}\,\D t\right]^{2(\nu-1)}.
\ee
Now the approximate pressure (\ref{pcha}) is given by 
\bn{pt}
p\approx\al\ga_m\left[-\frac{\sqrt{3}\,\al\ga_m}{2(\nu-1)}\,\D t\right]^{\nu-2},
\ee
while the approximate scale factors  is obtained by integrating the Friedmann equation (\ref{00}) as 
\bn{at}
a\approx a_s\left\{1-\frac{2(\nu-1)}{3\nu\al\ga_m}\left[-\frac{\sqrt{3}\,\al\ga_m}{2(\nu-1)}\,\D t\right]^\nu\right\}.
\ee
We introduce the main parameter
\bn{nu}
\nu=\frac{2(n+1)}{2n+1},
\ee
 which will be useful  for describing the different kinds of singularities in the near future (see Fig. 1). We  define  $\D t=t-t_s$   with  $t_s $ being  a finite cosmic  time.  Notice that the parameter $\nu$ vanishes for  $n=-1$ or diverges at  $n=-1/2$, hence, the expansions (\ref{rot})-(\ref{at}) are not well defined in those cases and we are going to deal with them separately. In fact,   we will be able to  solve the whole dynamics of the model and give the \emph{exact} scale factor  for both values of $\nu$. 

\begin{figure}
\begin{center}
\includegraphics[height=8cm,width=8cm]{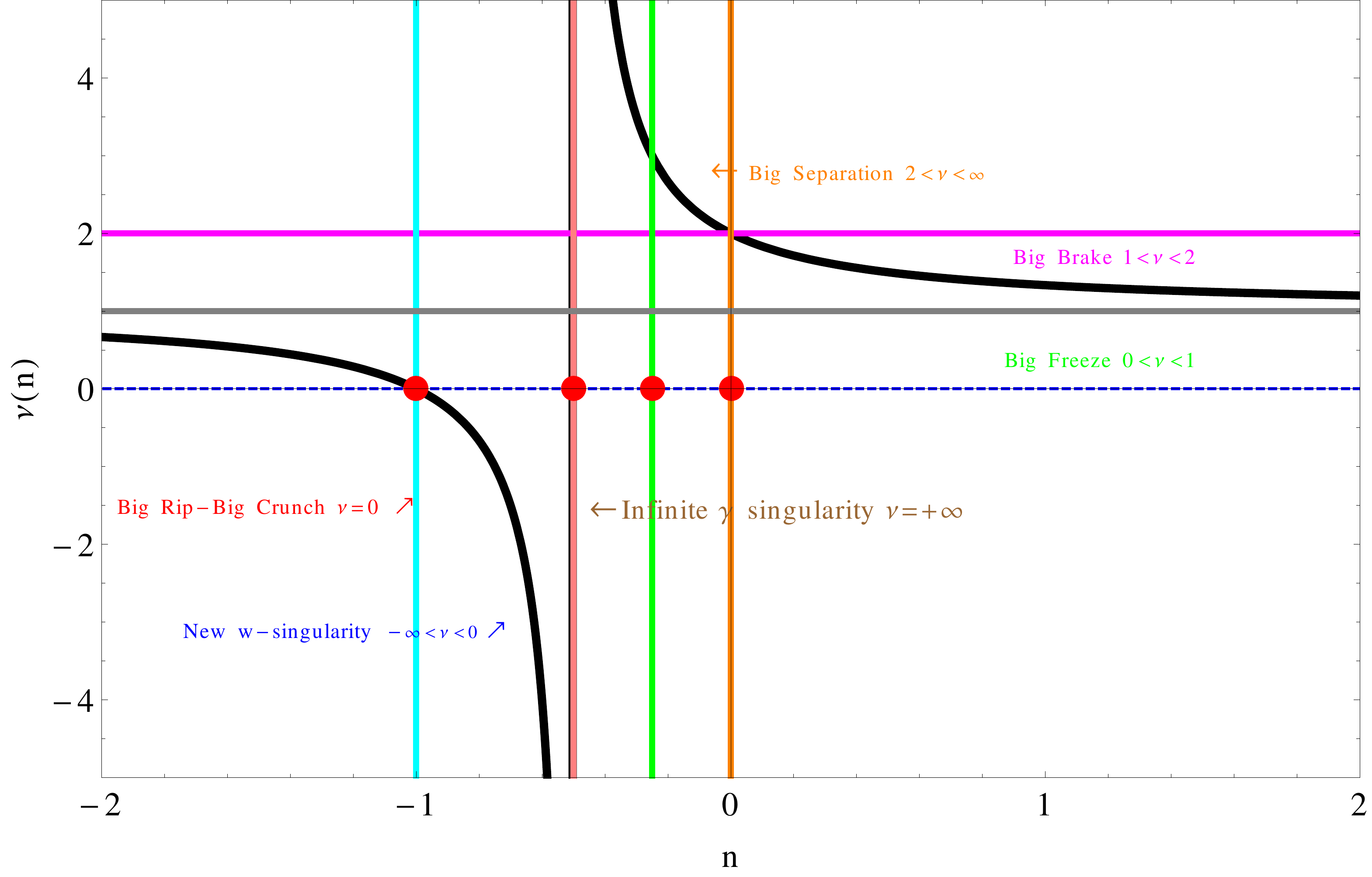}
\caption{Plot of  the main parameter $\nu$  in terms of $n$ indicating the range where the different kinds of singularities or abrupt events  are achieved.}
\label{si}
\end{center}
\end{figure}

Combining  the energy density (\ref{rot}),  (\ref{qf}),  and  (\ref{ro'f})-(\ref{rmxff}), we can rewrite  the interaction term, the approximate Hubble variable, its first, its second time derivatives, the acceleration term, the barotropic index,  and the matter and VVE densities, under the assumption that  the $\al\ro^{-n}$-term in Eq. (\ref{ro'f}) becomes dominant:  
\bn{qt}
Q\approx \frac{2(\nu-2)(\nu-1)}{3\ga_m\D t^2},
\ee
\bn{Ht}
H\approx-\frac{1}{\sqrt{3}}\left[-\frac{\sqrt{3}\,\al\ga_m}{2(\nu-1)}\,\D t\right]^{\nu-1},
\ee
\bn{H.t}
\dot H\approx-\frac{\al\ga_m}{2}\left[-\frac{\sqrt{3}\,\al\ga_m}{2(\nu-1)}\,\D t\right]^{\nu-2},
\ee
\bn{H..t}
\ddot H\approx\frac{\sqrt{3}\,\al^2\ga_m^2(\nu-2)}{4(\nu-1)}\left[-\frac{\sqrt{3}\,\al\ga_m}{2(\nu-1)}\,\D t\right]^{\nu-3},
\ee
\bn{act}
\frac{\ddot a}{a}\approx \dot H,
\ee
\bn{gat}
\ga\approx\frac{\al\ga_m}{\left[-\frac{\sqrt{3}\,\al\ga_m}{2(\nu-1)}\D t\right]^\nu},
\ee
\bn{rmxt}
\ro_m\approx \frac{-2\dot H}{\ga_m}, \qquad  \ro_x\approx \frac{2\dot H}{\ga_m}.  
\ee
We have carried out a detailed calculation of the  dynamical quantities (densities, pressures, interaction term, acceleration, etc) that we will employ for addressing the issue of cosmic singularities. The next step is to classify  the singularities using  the behavior of the above quantities near them.

%%%%%%%%%%%%%%%%%%%%%%%%%%%%%%%%%%%%%%%%%%%%%%%%%%%%%%%%%%%%%%%%%%%%%
\subsection{Classification of singularities}
%%%%%%%%%%%%%%%%%%%%%%%%%%%%%%%%%%%%%%%%%%%%%%%%%%%%%%%%%%%%%%%%%%%%%

After having discussed the main traits of an interacting cosmology   with two components  in the presence  of an initial or a final singularity, we must classify the different kinds of singularities that will emerge within this interacting framework. In order to do  that,  we  mainly use the approximate energy density (\ref{rot}), the pressure (\ref{pt}), the scale factor (\ref{at}) and remaining quantities (\ref{qt})-(\ref{rmxt}). We  classify the singularities according to the values taken by  the main parameter, $\nu$. Here, we will exclude the non interacting cases corresponding to $\nu=2$ and $\nu=1$. 

%%%%%%%%%%%%%%%%%%%%%%%%%%%%%%%%%
\vskip 0.5cm 
\no ${\it 1}$. ($1<\nu<2 $)-case {\bf Big Brake singularity}
\vskip 0.25cm 
%%%%%%%%%%%%%%%%%%%%%%%%%%%%%%%%%

\no If we now consider  the case of  a positive coupling constant ($\al>0$) and the  time approaching to the finite value $ t_s$ from the left $t\rightarrow t_s$ (thus $\D t<0$)  then  the scale factor of the universe (\ref{at}) reaches a finite value $a\to a_s$, the time derivative of the scale factor vanishes $\dot a\to 0$ but the second and third derivatives both diverge at the singularity ($\ddot a_s=-\infty$, $\dddot a_s=+\infty$). However, the energy density (\ref{rot}) along with  the Hubble variable $(\ref{Ht})$ are zero near the singularity, namely  $\ro(t_s)=0$ and $H(t_s)=0$. The time derivative of the Hubble variable and their subsequent ones diverge as well as the acceleration of the universe $\ddot{a}\approx a_s\dot H \to -\infty$, see Eq. (\ref{H.t}), while the pressure (\ref{pt}) positively grows without limit ($p\to+\infty$). Consequently, the interacting dark sector model has a late-time singularity at the finite cosmic  time $t_s$ for $n>0$ and $\al>0$ characterized by  a finite scale factor, a vanishing  time derivative,  energy density and  Hubble variable as well. But the acceleration and the pressure both diverge while the effective fluid fulfills anti-chaplygin gas equation of state (\ref{pcha}). These results show us that this kind of behavior corresponds to  a big Brake singularity, see \cite{tipo2a}. Interestingly enough,  it turned out the energy transfer from the VVE to the matter diverges as $t\to t_s$ ($Q\to-\infty$),   $\ro_m \rightarrow +\infty$,  and $\ro_x \rightarrow -\infty$   [see Eq.   (\ref{qt})].  It is important to stress that  the physical set up  associated with a big brake singularity   can be described in terms of a tachyonic scalar field \cite{tipo2a}.  A weaker extension of the big brake singularity  is obtained when a dust component is included in the Friedmann equation provided the Hubble function  does not vanish any longer at the singularity \cite{tipo2aa}.

We end this case  with the analysis of a complementary  branch where   the scale factor is  defined in the region  $t>t_s$  and  the coupling constant is negative ($\al<0$). The  former  branch which corresponds to  $\alpha>0$ and  $t<t_s$  matches with the latter one at $t=t_s$ up to its first time derivative whereas their higher order derivatives are all divergent \cite{tipo2aa}. Near the singularity the effects introduced by the interaction (\ref{qt}) on the matter and VVE components (\ref{rmxff}) are the same of  the $\al>0$ case, thus,  their limits are coincident.

%%%%%%%%%%%%%%%%%%%%%%%%%%%%%%%%%%%%
\vskip 0.5cm 
\no ${\it 2}$. ($2<\nu<\infty$)-case {\bf Big Separation singularity }
\vskip 0.25cm 
%%%%%%%%%%%%%%%%%%%%%%%%%%%%%%%%%%%%

In the interval of $\nu \in (2, \infty)$  under  the assumption of positive coupling constant ($\al>0$), we find that the scale factor (\ref{at}) reaches a constant value $a_s$ as $t\rightarrow t_s$ from the left   while the energy density (\ref{rot}) and the pressure (\ref{pt}) both approach to zero \cite{tipo2}, however,  the barotropic index (\ref{gat}) becomes divergent as $t\to t_s$. The first time derivative of the scale factor $\dot{a}\rightarrow 0$, its second derivative $\ddot{a}\rightarrow 0$, the Hubble variable $H\to 0$ and its first time derivative $\dot H\to 0$.  Taking into account that $\ro'=2\dot H\to 0$ as  $t\to t_s$ and $\ro''=2\ddot H/3H$, the source equation $(\ref{sef})$ then becomes 
\bn{H..}
\ddot H\approx\frac{3\ga_m}{2}H\,Q.
\ee
This equation explicitly shows that the second and subsequent time derivatives of the Hubble variable are strongly dependent on the interaction term. In fact, from Eqs. (\ref{H..t}) and (\ref{H..}), we find that  $H$, $\dot H$,  and their subsequently time derivatives $H^{(k)}$ vanish up to order $k\le [[\nu -1 ]]$ while the remaining ones diverge for $k\ge [[\nu -1 ]]$  if $t\to t_s$, where $[[x]]$ stands for the integer part of its argument. From Eqs. (\ref{qt}) and (\ref{rmxt}), we also have that $Q\to \infty$, $\ro_m \rightarrow 0$ and $\ro_x \rightarrow 0$  as  $t\rightarrow t_s$.  In conclusion,  the interacting dark sector model exhibits a big separation singularity, sometimes considered as a more softer singularity.

%%%%%%%%%%%%%%%%%%%%%%%%%%%%%
\vskip 0.5cm 
\no ${\it 3}$. ($\nu\to+\infty$)-case {\bf Exact solution} 
\vskip 0.25cm 
%%%%%%%%%%%%%%%%%%%%%%%%%%%%%

We will discuss in detail  the mysterious case associated with  $n\to-1/2$. In fact, we are going to show that the field equations can be solved exactly and therefore we can obtain a complete picture of  the  interacting  model. To this end, we introduce a new  variable $s=(a/a_s)^{3\ga_m/2}$ so that the Friedmann equation (\ref{00}) with the energy density  (\ref{rf}) reduces to the linear differential equation
\bn{s1/2}
\dot s=\pm\,\sqrt{3}\,\ga_m\al(1-s).
\ee
Having  solved this equation  we obtain four different types of solutions which need to be examined. Our starting point is to consider  the expanding  $a_1$-solution and its  effective barotropic index ($\ga=-2\dot H/3H^2$),
\bn{a1}
a_1=a_s\left[1+ {\mbox e}^{\frac{\sqrt{3}}{2}\al\ga_m \D t}\right]^{\frac{2}{3\ga_m}},
\ee
\bn{g1}
\ga_1=-\ga_m{\mbox e}^{-\frac{\sqrt{3}}{2}\al\ga_m \D t},
\ee
where  $H_1=\dot a_1/a_1 >0$ and $\ga_1<0$.  The contracting $a_3$-solution, obtained from the time reversal symmetry of the former one, leads to  $a_3=a_1(-t)$ with $H_3=\dot a_3/a_3 <0$ and $\ga_3=\ga_1(-t)$.  We name the another  expanding solution as  $a_4$  and is given by 
\bn{a4}
a_4=a_s\left[1- {\mbox e}^{-\frac{\sqrt{3}}{2}\al\ga_m \D t}\right]^{\frac{2}{3\ga_m}},
\ee
\bn{g4}
\ga_4=\ga_m{\mbox e}^{\frac{\sqrt{3}}{2}\al\ga_m \D t}.
\ee
Here  $H_4=\dot a_4/a_4>0$ and the contracting case is defined as   $a_2=a_4(-t)$ with $H_2=\dot a_2/a_2 <0$ and $\ga_2=\ga_4(-t)$. There we have assumed a positive coupling constant,  $\al>0$. The branches $a_1$ and $a_3$ are defined in the interval $-\infty<t<\infty$ while for the branches $a_2$ and $a_4$ are specified in the intervals  $t\le 0$ and $t\ge 0$, respectively. We have fixed the integration constants  so that the solution $a_1\to a_s$ and $a_3\to a_s$ in the limits $t\to-\infty$ and $t\to+\infty$, respectively.  We have fixed the final big crunch or the initial big bang singularity in $a_2$ or in $a_4$ at $t=0$. 

The solution $a_1$ ($a_3$) describes a universe which expands (contracts) from a finite scale factor $a_s$, free of an initial singularity, (from a infinite scale factor) at $t=-\infty$ and a final de Sitter stage (a finite scale factor). However, the solution $a_4$ ($a_2$) represents a universe which evolves from the big bang singularity (a finite scale factor $a_s$) at $t=0$ (in the limit $t=-\infty$) and increases (decreases) monotonically until reaches the finite scale factor $a_s$ (the big crunch singularity) in the limit $t\to\infty$ (at $t=0$). The last two scale factors $a_4$ and $a_2$ behave as a power-law solution $a\approx a_{s}[\sqrt{3}\al\ga_m\,|t|/{2}]^{2/3\ga_m}$ in the limit $|t|\to 0$ when the universe begins from a big bang singularity or ends in a big crunch singularity.

The scale factors $a_3$ and $a_4$ ($a_1$ and $a_2$) all of them go to $a_s$  as $\D t\to\infty$ ($\D t\to-\infty$) while its first and higher order time derivatives  vanish in those limits. Furthermore, we find that the Hubble variables can be written   in terms of the barotropic index as 
\bn{H}
H_i=\frac{\al}{\sqrt{3}}\frac{1}{\left(1-\frac{\ga_i}{\ga_m}\right)},
\ee
where the index $i$ runs from $1$ to $4$.  Their subsequent time derivatives $H_i^{(k)}\to 0$ as $a\to a_s$.

The energy density of the mix, its pressure and the interaction term  read 
\bn{r12}
\ro_i=\al^2\left[-1+\left(\frac{a_s}{a_i}\right)^{\frac{3\ga_m}{2}}\right]^{2},
\ee
\bn{p12}
p_i=(\ga_m-1)\ro_i+\al\ga_m\ro_i^{1/2},
\ee
\bn{q12}
Q_i=\frac{\al\ga_m}{2}\left[\al+\ro_i^{1/2}\right].
\ee
At this point,  we highlight that the time dependence of these quantities can be obtained by replacing  each one of the exact scale factors found before [see Eqs. (\ref{a1}) and (\ref{a1})]. From (\ref{r12}) and (\ref{p12}), we have that the energy density $\ro(a)\to 0$  and the pressure $p\to 0$ in the limit $a\to a_s$. Although the energy density and the pressure vanish  in this limit, however, the effective barotropic index $\ga=(\ro+p)/\ro=\ga_m[1+\al\ro^{-1/2}]$ diverges. Most importantly,  the higher order time derivatives of $\ga$ also diverge as $a\to a_s$  so we name this behavior as ``infinite $\ga$-singularity".

Consequently, the universe presents an infinite  $\ga$-singularity  in the distant past or the remote future ($t=\pm\infty$) where the scale factor reaches a finite values $a_s$, but  the first and subsequent time derivatives of the scale factor vanish, further,  the Hubble variable along with their higher order time derivatives also vanish near $a_s$. We remark  this ``new singularity'' differs from the one reported in \cite{wsingu1} provided it could not be  reached at finite time. Nevertheless,  this singularity is   characterized by a non-vanishing interaction term (\ref{q12})  at $a_s$, $Q\to \al^2\ga_m/2$,   while the matter component $\ro_m\to 0$ and the VEE $\ro_x\to 0$. Even though all physical quantities ($p$, $\rho$, $\ro_m$, $\ro_x$ )  fades away quickly  as one approaches to the singularity, in the far remote future or distant asymptotic past,  the interaction cannot be turned off neither the barotropic index.  % $a\to a_s$ 

%%%%%%%%%%%%%%%%%%%%%%%%%%%%%%%%%%%%
\vskip 0.5cm 
\no ${\it 4}$. ($-\infty<\nu<0$)-case {\bf New $w$-singularity} 
\vskip 0.25cm
%%%%%%%%%%%%%%%%%%%%%%%%%%%%%%%%%%%%
In addition to the spacetime singularities at finite  time mentioned above there are another kinds of abrupt events \cite{mariam3} related with the ultimate fate of the universe which indeed requires some mentioning for sake of completeness. The abrupt events were discovered within the framework of dark energy model and attracted some attention  mostly beacuse are less dangerous than  singularity. Concerning this aim, we are going to analyze a new kind of abrupt event that is related with the so called $w$-singularity found in \cite{wsingu1} and re-examined in \cite{wsingu2}.

For these values of the main parameter $\nu$, the scale factor (\ref{at}), the energy density (\ref{rot}) and the pressure (\ref{pt}) have the limits $a\to a_s$, $\ro\to 0$ and $p\to 0$ as the time variable $t\to\infty$ while the first and higher order time derivatives of these three quantities vanish  in the remote future($t\to\infty$).  Due to that the effective barotropic index (\ref{gat}) and their subsequent  time derivative $ \ga^{(k)}$ diverges up to order $k\le [[ -\nu]]$ but vanishes for all the following ones in the limit  $t\to\infty$; as a result  the universe ends in an abrupt $\ga$-singularity  in the remote future, again, this kind of singular event differs from the case explored in \cite{wsingu1}. The approximate interaction term (\ref{qt}) vanishes in the abrupt event whereas Eq. (\ref{rmxt}) shows us that  $\ro_x \rightarrow 0$ and $\ro_m \rightarrow 0$  as $a\rightarrow a_s$.  This singular event differs from the  ``infinite $\ga$-singularity"  because  the interaction  vanishes in the latter case while the former one  leads to $Q\to \al^2\ga_m/2$.

%%%%%%%%%%%%%%%%%%%%%%%%%%%%%%%%%
\vskip 0.5cm 
\no ${\it 5}$. ($\nu=0$)-case {\bf Big Rip/Crunch Singularity} 
\vskip 0.25cm
%%%%%%%%%%%%%%%%%%%%%%%%%%%%%%% %%

%\cite{tipo1}, \cite{tipo1b}.
For this particular value of the main parameter $\nu$, the constant $n=-1$ and the interaction term (\ref{q1}) becomes a linear function of the $\eta$-derivative of the energy density, $Q=-\al\ro'$, and  therefore the first integral (\ref{ro'1}) turns to be 
\bn{-1}
\ro'+\ga_{m}(1+\al)\ro=c\ga_m.
\ee
We will solve this equation for the energy density and below show the general exact solution of the Friedmann equation for any value of the integration constant $c$. This includes the corresponding power law  scale factor for $c=0$: 
\bn{a-1c0}
a=a_st^{2/3\ga_m(1+\al)},   \qquad  c=0,
\ee
\bn{r-1c0}
\ro=\frac{4}{3\ga_m^2(1+\al)^2}\left(\frac{a_s}{a}\right)^{3\ga_m(1+\al)},   \qquad  c=0,
\ee
and two families of solutions associated with  $c\ne 0$,
\bn{a-1}
a=a_s\left[\frac{1-\cos{\om t}}{2}\right]^{1/3\ga_m(1+\al)}, 
\ee
\bn{r-1}
\ro=\frac{3c^2\ga_m^2}{\om^2}\left[\frac{\sin{\om t}}{1-\cos{\om t}}\right]^{2},
\ee
where $\quad \om^2=3c\,\ga_m^2 (1+\al)>0$ and
\bn{a-1-}
a=a_s\left[\frac{1+\cosh{\om t}}{2}\right]^{1/3\ga_m(1+\al)}, 
\ee
\bn{r-1-}
\ro=\frac{3c^2\ga_m^2}{\om^2}\left[\frac{\sinh{\om t}}{1+\cos{\om t}}\right]^{2},
\ee
where $\quad \om^2=-3c\,\ga_m^2 (1+\al)>0$ while the pressure and the interaction term are
\bn{p-1}
p=c\ga_m+[\ga_m(1+\al)-1]\ro,
\ee
\bn{q-1}
Q=\al\ga_m[c+(1+\al)\ro].
\ee

For $\om^2=3c\,\ga_m^2 (1+\al)>0$ and  $\al>-1$, we have that the integration constant $c>0$ and the solution (\ref{a-1}) represents a universe with a finite time span that begins with a big bang at $t=0$, then passes by a maximum at $t_{\rm max}=\pi/\om$ and ends in a big crunch at $t_{\rm bc}=2\pi/\om$ \cite{tipo2}. However,  for $\al<-1$ and $c<0$ there is a significant difference in the behavior of the  solution  (\ref{a-1}) provided  the  universe has a finite time span but begins with a contracting phase at $t=0$ associated with an infinite scale factor, then bounces at $t_{\rm bounce}=\pi/\om$  and ends with an infinite scale factor in a big rip singularity at $t_{\rm br}=2\pi/\om$ \cite{tipo2}. Near the big bang, big crunch and big rip singularities, the energy density (\ref{r-1}), the pressure (\ref{p-1}), interaction term (\ref{q-1}) and the matter and the VVE densities  (\ref{rmxt}) diverge, namely $\ro\to\infty$, $p\to-\infty$, $Q\to+\infty$, $\ro_m\to-\infty$, and $\ro_x\to+\infty$. At the extrema of the scale factor (\ref{a-1}) $t_{\rm max/bounce}=\pi/\om$, we have that $\ro$ vanishes, both $p$ and $Q$ become constants $p=c\ga_m$, $Q=\al p$  while $\ro_m=\ro_x=0$. 

For $\om^2=-3c\,\ga_m^2 (1+\al)>0$ and $\al>-1$, we have that the integration constant $c<0$ and the solution (\ref{a-1-}) describes a universe with an initial contracting de Sitter phase in the remote past, then bounces at $t=0$ and ends in an expanding de Sitter stage in the remote future. For $\al<-1$ and $c>0$, the universe evolves from a zero radius (vanishing scale factor) at the distant past, then passes by a maximum at $t=0$ and ends in the remote future, again,  with a vanishing scale factor. The energy density (\ref{r-1-}), pressure (\ref{p-1}), the interaction term (\ref{q-1}) and the matter  and VVE densities (\ref{rmxff}) approach to the constants $\ro\to 3c^2\ga_m^2/\om^2$, $p\to-3c^2\ga_m^2/\om^2=-\ro$, $Q\to\al\ga_m(1-c)$, $\ro_m\to-c$ and $\ro_x\to c\al/(1+\al)$ as time $t\to\pm\infty$. At $t=0$ the scale factor exhibits  an extreme where   $\ro=0$, $p=c\ga_m$, $Q=\al p$ and $\ro_m=\ro_x=0$. 

\vskip .5cm

%%%%%%%%%%%%%%%%%%%%%%%%%%%%%%%%%%%%%%%%%%%%
\vskip 0.5cm 
\no ${\it 6}$. ($0<\nu<1$)-case {\bf Big Freeze} 
\vskip 0.25cm 
%%%%%%%%%%%%%%%%%%%%%%%%%%%%%%%%%%%%%%%%%%%%

The big freeze singularity was firstly encountered in the literature by Bouhmadi-L\'opez \emph{et al.} within the context of a phantom generalized Chaplygin gas \cite{mariam}. Further, they proved  that this singularity at a finite scale factor arises  in a Randall-Sundrum I brane-world scenario if the brane is filled with the dual version of the generalized phantom Chaplygin gas \cite{mariam}. Besides, the avoidance  of this singularity within the context of  quantum cosmology was explored  with the help of  the Wheeler-de Witt equation by  mimicking a (phantom) generalized Chaplygin gas with a  scalar field \cite{mariam2}. 

The scale factor (\ref{at}) approaches to a finite value $a\to a_s$ in the limit $t\rightarrow t_s$. However, its  time derivative $\dot a$ and their subsequent ones $a^{(k)}$ with $k>1$ diverge for $t\to t_s$, indicating the existence of the big freeze singularity \cite{tipo3}. Concerning  the Hubble variable (\ref{Ht}), the energy density (\ref{rot}), the pressure (\ref{pt}), the barotropic index (\ref {gat}) and the interaction term (\ref{qt}), these and their higher order time derivatives diverge in the limit $t\to t_s$. Then, we have a divergent behavior of the dark matter and VVE density as $t\to t_s$.  

%%%%%%%%%%%%%%%%%%%%%%%%%%%%%%%%%%%%%%%%%%%%%%%%%%%%%%%%%%%%%%%%%%%%%
\subsection{Krolak and Tipler criteria}
%%%%%%%%%%%%%%%%%%%%%%%%%%%%%%%%%%%%%%%%%%%%%%%%%%%%%%%%%%%%%%%%%%%%%
%%PONER los invariantes de curvaturas aqui!!!!!!!!

In what follows, we give a reinterpretation of the above classification making a description of the singularities and/or abrupt events from a geometric point of view  based on a method developed by Tipler \cite{tipler} and Kr\'olak \cite{krolak}.  A spacetime is Tipler strong \cite{tipler}  iff as the proper  time   $t\rightarrow t_{s}$, the integral 
\bn{ti}
{\cal {T}}(t)=\int^{t}_{0}dt'\int^{t'}_{0}{|{\cal{R}}_{ab}u^{a}u^{b}|}dt'' \rightarrow \infty.
\ee
In same manner,  a spacetime is Kr\'olak strong \cite{krolak}  iff as the proper  time  $t\rightarrow t_{s}$, the integral 
\bn{ki}
{\cal {K}}(t)=\int^{t}_{0}{|{\cal{R}}_{ab}u^{a}u^{b}|}dt \rightarrow \infty,
\ee
where the component of Ricci tensor are  understood to be written in a  parallel transported frame along the geodesic curves.  Notice that a singularity can be strong by Kr\'olak criteria but weak according to Tipler's criteria, however, the reverse situation always holds. Because weak singularities can be extended beyond them, the method developed by Tipler and Kr\'olak are  useful tools for determining the fate of the universe in terms of the fate of geodesic curves near potential strong singular point.

 Let us consider  time-like geodesic curves, $x^{i}=c$ with $i$ spatial index and $c$ a constant \cite{geo}, associated with co-moving observer, i.e,  we take into account a co-moving world-line congruence with velocity $u^{\alpha}=(\partial_{t})^{a}=(1, 0,0,0)$ so that the proper time and the coordinate time are the same. Moreover, the components of the Ricci tensor measured by an observer along this congruence  lead to ${\cal{R}}_{~ab}u^{a}u^{b}=-3\ddot{a}(t)/a(t)$. Using the latter fact together  (\ref{at}) and (\ref{H.t}) is not difficult to see  that a big brake singularity is T-weak and K-weak provided both integrals  (\ref{ti})-(\ref{ki})  vanish for $\nu \in (1,2)$, namely ${\cal {T}} \propto \D\tau^{\nu}$ and ${\cal {K}} \propto \D\tau^{\nu-1}$ as $\D\tau \rightarrow 0$. Nevertheless, the leading term in the square Riemann ${\cal R}^{abcd} {\cal R}_{abcd} \propto (\ddot{a}/a)^2$, given by    $\D\tau^{2(\nu-2)}$,  diverges as  $\D\tau \rightarrow 0$ \cite{inva}. Also the Ricci scalar, ${\cal R}\propto  \ddot{a}/a \propto \D\tau^{\nu-2}$, blows up at the singularity \cite{inva}. A big separation singularity can be considered as a weaker event provided  is not only T-weak and K-weak but also  ${\cal R}^{abcd} {\cal R}_{abcd}$  and ${\cal R}$ both vanish at the singularity. The ``new $w$-singularity'' is T/K-weak and avoids divergences in geometric invariant as ${\cal R}^{abcd} {\cal R}_{abcd}$ in the asymptotic past or remote future, indeed   a similar situation occurs for the infinite $\gamma$-singularity.  On the other hand, the behavior of scale factor near a big rip singularity is $a(t) \propto (t-t_{\rm {br}})^{-p}$ with $p>0$,  so we classify it as a K/T-strong singularity, and all scalar invariants blow up at $t_{\rm {br}}$. Besides, one could expect that both  big bang and big crunch singularity exhibit a similar behavior regarding the K/T criteria or the blow up of ${\cal R}^{abcd} {\cal R}_{abcd}$. In the case of a big freeze ultimate fate,  the leading term in Krolak strength gives ${\cal {K}} \propto \D\tau ^{\nu-1}$ so it diverges as $t\rightarrow t_{s}$, however, Tipler measure involves a second integration and yields ${\cal {T}} \propto (t -t_{s})^{\nu}$,  being totally regular as $t\rightarrow t_{s}$ provided $\nu \in (0,1)$. Notice that the squared Riemann and the Ricci scalar both diverge for the big freeze event, as can be seen from  (\ref{at}) and (\ref{H.t}). It is important to stress that the analysis performed with casual geodesic  which meets  a $T/K$-strong singularity  gave the same kind of finding because involves the  integral  of component  Riemann tensor parallel transported, ${\cal{R}}^{a}_{~bcd}u^{b}u^{d}$, for which the non-vanishing components turned to be the same ($ {\cal{R}}^{a}_{~tct}=-\ddot{a}/a$) also \cite{ruth}.   We end this section by studying the violation or not of the energy conditions for the aforesaid singularities \cite{ec}. Interestingly enough,  we obtained that all the energy conditions (WEC, NEC, SEC) are fulfilled near the singularities except for  a few cases, namely  the big rip and  the infinite $\gamma$ singularity (associated with the expanding solution $a_{1}$) both violate SEC \cite{ec}.    
 
An open question related with the present work is  what kind of fundamental theory can support these singularities? In this direction, Barrow and Graham have recently shown that a new ultra-weak generalized sudden singularity can be supported by a scalar field  with a simple power-law potential \cite{barrowfinale}. We hope to return to this issue in the near future. 
 
  %The big brake, big separation, infinite $\gamma$-event, new $w$-singularity, and big freeze singularity do not brake  the null, weak or strong energy conditions. On the other hand, a big rip singularity violates all the three  energy conditions mentioned. 

%REVISAR el caso big freez-EC!!!!!
%%%%%%%%%%%%%%%%%%%%%%%%%%%%%
%%%%%%%%%%%%%%%%%%%%%%%%%%%%%%%%%%
\section{Conclusions}
%%%%%%%%%%%%%%%%%%%%%%%%%%%%%%%%%%
We have investigated the evolution of an interacting dark sector model in a  spatially flat FRW spacetime, solved exactly the source equation when the energy exchange is produced by a nonlinear interaction term and found that the equation of state of the effective dark fluid turned to be that of  a Chaplygin or anti-Chaplygin gas according to the sign of the coupling constant. We highlighted that this interacting dark matter-VVE model, generated by a nonlinear interaction term, harbored  future-like singularities related with the ultimate fate of the universe and two abrupt events called  infinite $\gamma$ singularity and new $w$-singularity. 

We have obtained the approximate scale factor in the vicinity of the singular events and used it to study physically viable models describing the current along with final state of the universe, in particular, we have examined the behavior of the interaction and  the dark component energy densities near the singularities.  We have classified entirely the singularities of the model for any value of the main parameter $\nu$ while for certain value of $\nu$, we have shown that the two-fluid system is fully integrable, so the exact form of the scale factor and the remaining representative quantities of the model were found.   

Using the aforesaid scale factor and the geometric method developed by Tipler and Kr\'olak for the case of time-like geodesic curves, associated with co-moving observer, we have obtained a regular behavior of Tipler measure in the case of big separation, big brake,  big freeze,  new $w$-singularity, and the infinite $\gamma$-singularity,  making them traversable in the sense that a time-like geodesic curves can be extended beyond such singular events. However, big freeze singularities are K-strong due to Kr\'olak measure becomes divergent  near them. On the other side, we have shown that  brig rip singularity cannot be smoothed within interacting scenarios and therefore Tipler and Kr\'olak measures diverges as geodesics meet such final event. 

%%%%%%%%%%%%%%%%%%%%%%%%%%%%%%%%%%%%%%%%%%%%%%%%%%
\acknowledgments
%%%%%%%%%%%%%%%%%%%%%%%%%%%%%%%%%%%%%%%%%%%%%%%%%

L.P.C thanks  U.B.A under Project No. 20020100100147 and CONICET under Project PIP 114-201101-00317.  M.G.R  is partially supported by CONICET. 

%%%%%%%%%%%%%%%%%%%%%%%%%%%%%%%%%%%%%%%%%%%%%%%


\begin{thebibliography}{99}
%%%%%%%%%%%%%%%%%%%%%%%%%%%%%%%%%%%%%%%%%%%%%%%

\bibitem{Book}
Y. Wang, ``Dark energy'', Wiley-vch Verlag GmbH and Co. KGaA, ISBN 978-527-40941-9 (2010);
``Dark energy: Observational and theoretical approaches'', edited by Pilar Ruiz-Lapuente, Cambridge University Press 2010.
\bibitem{Planck2013}
%Planck 2013 results. XVI. Cosmological parameters
P. A. R. Ade \emph{et al}, [arXiv:1303.5076v1].
\bibitem{WMAP9}
%NINE-YEAR WILKINSON MICROWAVE ANISOTROPY PROBE (WMAP) OBSERVATIONS:COSMOLOGICAL PARAMETER RESULTS
G. Hinshaw \emph{et al.}, [arXiv:1212.5226v3].

\bibitem{tipo1}
R. R. Caldwell, Phys. Lett. B {\bf 545}, 23 (2002).
\bibitem{tipo1b}
A. A. Starobinsky, Grav. Cosmol. {\bf 6}, 157 (2000);
L. P. Chimento and R. Lazkoz, Phys. Rev. Lett. {\bf 91},
211301 (2003);
M. P. D¸abrowski, T. Stachowiak and M. Szyd lowski,
Phys. Rev. D {\bf 68}, 103519 (2003);
 P. F. Gonz\'alez--D\'iaz, Phys. Rev. D {\bf 69}, 063522 (2004);
S. i. Nojiri and S. D. Odintsov, Phys. Rev. D {\bf 70}, 103522
(2004);
A. Balcerzak and
M. P. Dabrowski, Phys. Rev. D {\bf 73} (2006) 101301.
\bibitem{tipo2}
S. i. Nojiri, S. D. Odintsov and S. Tsujikawa, Phys. Rev.
D {\bf 71}, 063004 (2005); M.P. Dabrowski, K. Marosek, and A. Balcerzak, [arXiv: 1308.5462].
\bibitem{tipo2a}
%%bigbrake
 V. Gorini, A. Y. Kamenshchik, U. Moschella and
V. Pasquier, Phys. Rev. D {\bf 69}, 123512 (2004).
\bibitem{tipo2aa}
%Paradox of soft singularity crossing and its resolution by
%distributional cosmological quantities, Phys. Rev. D 86,
063522 (2012
%Extension of BBrake singularity
Z. Keresztes, L.A. Gergely,  and  A. Y. Kamenshchik, Phys. Rev. D {\bf 86}, 063522 (2012).  
\bibitem{tipo2b}
J. D. Barrow, Class. Quant. Grav. {\bf 21}, L79 (2004);
J. D. Barrow
and C. G. Tsagas, Class. Quant. Grav. {\bf 22}
(2005) 1563.

\bibitem{tipo3}
%review
S. ’i. Nojiri and S. D. Odintsov, Phys. Rev. D {\bf 72}, 023003
(2005); A. Yu. Kamenshchik, Class. Quantum Grav. 30, 173001
(2013) [arXiv:1307.5623v2].

\bibitem{mariam}
%big freeze
 M. Bouhmadi-L\'opez, P. F. Gonz\'alez-D\'iaz and P. Mart\'in-
Moruno, Phys. Lett. B 659, 1 (2008);
 M. Bouhmadi-L\'opez, P. F. Gonz\'alez-D\'iaz and P. Mart\'in-
Moruno, Int. J. Mod. Phys. D {\bf 17}, 2269 (2008).
\bibitem{mariam2}
Mariam Bouhmadi-L\'opez, Claus Kiefer, Barbara Sandhofer, Paulo Vargas Moniz, [arXiv:1002.4783].
\bibitem{tipo4}
S. ’i. Nojiri and S. D. Odintsov, Phys. Rev. D {\bf 78}, 046006
(2008);
 K. Bamba, S. ’i. Nojiri and S. D. Odintsov, JCAP {\bf 0810},
045 (2008).
\bibitem{haw}
S.W. Hawking, G.F.R. Ellis, The Large Scale Structure
of Space-time, Cambridge University Press, Cambridge,
(1973).


\bibitem{ruth}
%Classification of cosmological milestones
L. Fernandez-Jambrina, R. Lazkoz, Phys.Rev. D {\bf 74} (2006) 064030;
%Geodesic behavior of sudden future singularities
L. Fernández-Jambrina, Ruth Lazkoz, Phys.Rev. D {\bf 70} (2004) 121503.
\bibitem{barrow}
%Geodesics at Sudden Singularities
John D. Barrow, S. Cotsakis, Phys. Rev. D {\bf 88} (2013) 067301.


\bibitem{tipler}
F.J. Tipler, Phys. Lett. A {\bf 64} (1977) 8.
\bibitem{krolak} 
A. Krolak, Class. Quant. Grav. {\bf 3} (1986) 267;
C.J.S. Clarke and A. Kr´olak, Journ. Geom. Phys. {\bf 2}
(1985) 17.
\bibitem{k2}
A. Kr\'olak and W. Rudnicki, Inter. J. Theor. Phys. {\bf 32} (1993), 137-142.
\bibitem{jefe1}
L.P.Chimento, Phys.Rev.D{\bf81} 043525 (2010).
\bibitem{jefe2}
L. P. Chimento, M. G. Richarte, Phys.Rev. D {\bf 84} 123507 (2011);
L. P. Chimento, M. G. Richarte, Phys.Rev. D {\bf 85}  127301 (2012);
%Dark matter, dark energy, and dark radiation coupled with a transversal interaction
L. P. Chimento and M. G. Richarte, Phys. Rev. D {\bf 86} 103501 (2012);
%Dark radiation and dark matter coupled to holographic Ricci dark energy
L. P. Chimento, M. G. Richarte, Eur.Phys.J. C {\bf 73} (2013) 2352;
L. P. Chimento, M. G. Richarte, Eur.Phys.J. C {\bf 73} (2013) 2497;
%Interacting dark sector with variable vacuum energy
L. P. Chimento, M. G. Richarte, I. E. S. Garc\'ia, Phys. Rev. D {\bf 88} 087301 (2013), [arXiv:1310.5335]. 
%\bibitem{demo}
%It can be shown that the interaction proposed here is transparent to this condition, so it  did not lose generality with such assumption.
\bibitem{mariam3}
%Tradeoff between Smoother and Sooner "Little Rip" (classification)
Mariam Bouhmadi-Lopez, Pisin Chen, Yen-Wei Liu, [arXiv:1302.6249];
%Grand Rip-GrandBang-classification
L. Fern\'andez-Jambrina, [arXiv:1408.6997];
%T. Ruzmaikina and A. A. Ruzmaikin, Sov. Phys. JETP30, 372 (1970);
H. Stefancic, Phys. Rev. D  {\bf 71}, 084024 (2005); 
M. Bouhmadi-L\'opez, Nucl. Phys. B {\bf 797}, 78 (2008).

\bibitem{wsingu1}
M. P. Dabrowski and T. Denkieiwcz, Phys. Rev. D {\bf 79},
063521 (2009).

\bibitem{wsingu2} 
L. Fern\'andez-Jambrina,  Phys.Rev.D {\bf 82} 124004  (2010).
\bibitem{geo}
In a FRW spacetime, curves of the form $x^{i}(t)=c$ with $t$ the proper time fulfill the geodesic equation, $\ddot{x}^{\nu}+\Gamma^{\nu}_{~\al\nu}u^{\al}u^{\nu}=0$, provided the only component of 4-velocity is $u^{t}=1$, $\dot{u^{t}}$=0,  and  $\Gamma^{t}_{~tt}=0$.
\bibitem{inva}
For a FRW metric,   the squared  Riemann is given by ${\cal R}^{abcd} {\cal R}_{abcd}=12 \left(\frac{\dot{ a}^4}{a^4}+ \frac{\ddot{a}^2}{a^2}\right)$ and the Ricci scalar is  ${\cal R}=6(H^2+ \frac{\ddot{a}}{a})$.
\bibitem{ec}
The weak energy condition corresponds to the case with $\rho\ge0$ and $\rho + p \ge 0$, the null condition only involves the latter inequality ($\rho + p \ge 0$) whereas the strong energy condition is satisfied if $\rho + p \ge0$ and $\rho + 3p \ge 0$. 
\bibitem{barrowfinale}
J. D. Barrow and A. H. Graham, [arXiv:1501.04090v2].
\end{thebibliography}
\end{document}